\documentclass[journal,draftclsnofoot,onecolumn,12pt]{IEEEtran}

\usepackage[T1]{fontenc}% optional T1 font encoding

\usepackage{cite}
\usepackage[colorlinks,
linkcolor=red,
anchorcolor=blue,
citecolor=green
]{hyperref}
\ifCLASSINFOpdf
\else
\fi
\usepackage{amsmath}
\usepackage{diagbox}

\usepackage{ amssymb }
\usepackage{amsfonts}
\usepackage{caption}
\usepackage{graphicx}
\usepackage{lettrine}
\usepackage{epstopdf}
\usepackage{textcomp,booktabs}
\usepackage[usenames,dvipsnames]{color}
\usepackage{colortbl}
\definecolor{mygray}{gray}{.9}
\definecolor{mypink}{rgb}{.99,.91,.95}
\definecolor{mycyan}{cmyk}{.3,0,0,0}
\usepackage{pifont}
\usepackage{subfig}
\usepackage{multirow}
\usepackage{url}
\usepackage{color}
\usepackage{threeparttable}
\usepackage{setspace}
\usepackage{lineno}

\usepackage[dvipsnames,usenames]{color}

\usepackage{bm}

\usepackage{algorithm} %format of the algorithm
\usepackage{algorithmic} %format of the algorithm

\usepackage{dblfloatfix}

\usepackage[dvipsnames,usenames]{color}
\usepackage[usenames,dvipsnames]{color}

\definecolor{light-gray}{gray}{0.90}
\begin{document}
%\linenumbers
%\pagestyle{empty}  % no page number for the second and the later pages
%\topmargin=-0.6 truein
%\leftmargin=-0.6 truein
%\rightmargin=-0.6 truein
%\textheight=9.15 truein
%\oddsidemargin=-0.15truein
%\evensidemargin=-0.15truein
%\textwidth=6.5truein
	\title{Wireless Semantic Transmission via  Revising \\ Modules in Conventional Communications}

	\author{Peiwen Jiang, Chao-Kai Wen, Shi Jin, and Geoffrey Ye Li
			\thanks{P. Jiang and S. Jin are with the National
				Mobile Communications Research Laboratory, Southeast University, Nanjing
				210096, China (e-mail: PeiwenJiang@seu.edu.cn; jinshi@seu.edu.cn).}
			\thanks{C.-K. Wen is with the Institute of Communications Engineering, National
				Sun Yat-sen University, Kaohsiung 80424, Taiwan (e-mail: chaokai.wen@mail.nsysu.edu.tw).}
			\thanks{G. Y. Li is with the Department of Electrical and Electronic Engineering,
				Imperial Colledge London, London, UK (e-mail: geoffrey.li@imperial.ac.uk).}}
	
	\maketitle
	\pagestyle{empty}  % no page number for the second and the later pages
	\thispagestyle{empty} % no page number for the first page
	%\vspace{-0.5cm}
	%\begin{abstract}
%		
%		
%		\setlength{\belowcaptionskip}{-0.3cm}   %调整图片标题与下文距离
%		\begin{figure*}[!t]
%			\centering
%			\includegraphics[width=6in]{gaoxuanxuan2}
%	
\begin{abstract}
Semantic communication has become a popular research area due its high spectrum efficiency and error-correction performance.  Some studies use deep learning to extract semantic features, which usually form end-to-end semantic communication systems and are hard to address the varying wireless environments. Therefore, the novel semantic-based coding methods and performance metrics have been investigated and the designed semantic systems consist of 
various modules as in the conventional communications but with improved functions.  This article discusses recent achievements in the state-of-art semantic communications exploiting the conventional modules in wireless systems.  We demonstrate through two examples that  the traditional hybrid automatic repeat request and modulation methods can be redesigned for novel semantic coding and metrics to further improve the performance of  wireless semantic communications. At the end of this article, some open issues are identified.

\end{abstract}
	
%	\newpage
	
%	\vspace{0.2cm}
%	\renewcommand{\baselinestretch}{1.5}
%	\setlength{\baselineskip}{22pt}
	\section{Introduction}

 	\IEEEPARstart{S}{emantic} communication can  significantly reduce the requirements of transmission resources. Different from conventional communications, semantic-based methods commonly rely on a knowledge base (KB)  to remove redundancy  and correct errors during  transmission. The KB can be represented by a specific content or a set of trainable nerual networks. Compared to  bit-level transmission in the conventional communications, semantic communications are content-related and  directly transmits the desired meaning. %Owing to development of deep learning (DL), semantic communication is easily  established while KB is difficultly extracted manually.
 	 Deep learning (DL)-based semantic communication usually  realizes a content-related coding and decoding based on the same KB in an end-to-end (E2E)  manner. A proper KB  is essential to the spectrum efficiency and error-correction performance of a semantic system. Therefore, domain adaption \cite{zhang2022unified} and federated learning \cite{tong2021federated} methods have been developed to update and share a new KB for both the transmitter and the receiver.  
 	
 	Recently, semantic communication is implemented by redesigning or revising the modules in the conventional communications, which can be better address varying wireless environments. In addition to semantic coding and decoding, other modules in the conventional communication systems also need to be  adjusted due to the change of transmission contents from symbol sequence to semantic meaning and performance metrics.  For semantic communications, the  modulation method is redesigned in \cite{guo2022signal} to maximize the sentence similarity rather than to minimize the bit errors. This metric change significantly affects the  modulation because the words with similar meaning can be modulated into close constellation points.  In \cite{shao2022semantic}, peak-to-average-power ratio (PAPR)  is also reduced with semantic coding together  to improve the semantic similarity between the received and transmit sentences. Hybrid automatic repeat request (HARQ) is  a key technique to address varying wirelss channels in the conventional communications, which has been exploited to develop the semantic-based HARQ in \cite{jiang2022deep}. Furthermore, the semantic transmitter can  adaptively carry different amounts of semantic contents   according to the channel information \cite{zhou2022adaptive}.    To accommodate different performance metrics and requirements, the resource allocation \cite{mu2022heterogeneous} for multi-user wireless communications becomes heterogeneous; therefore, the complexity is sharply increased. In general, the novel performance metrics and transmission methods for semantic communications require a brand-new design for wireless communications.
 	
\begin{figure*}[!h]
	\centering

	%	\subfloat[ ]{
	\includegraphics[width=7in]{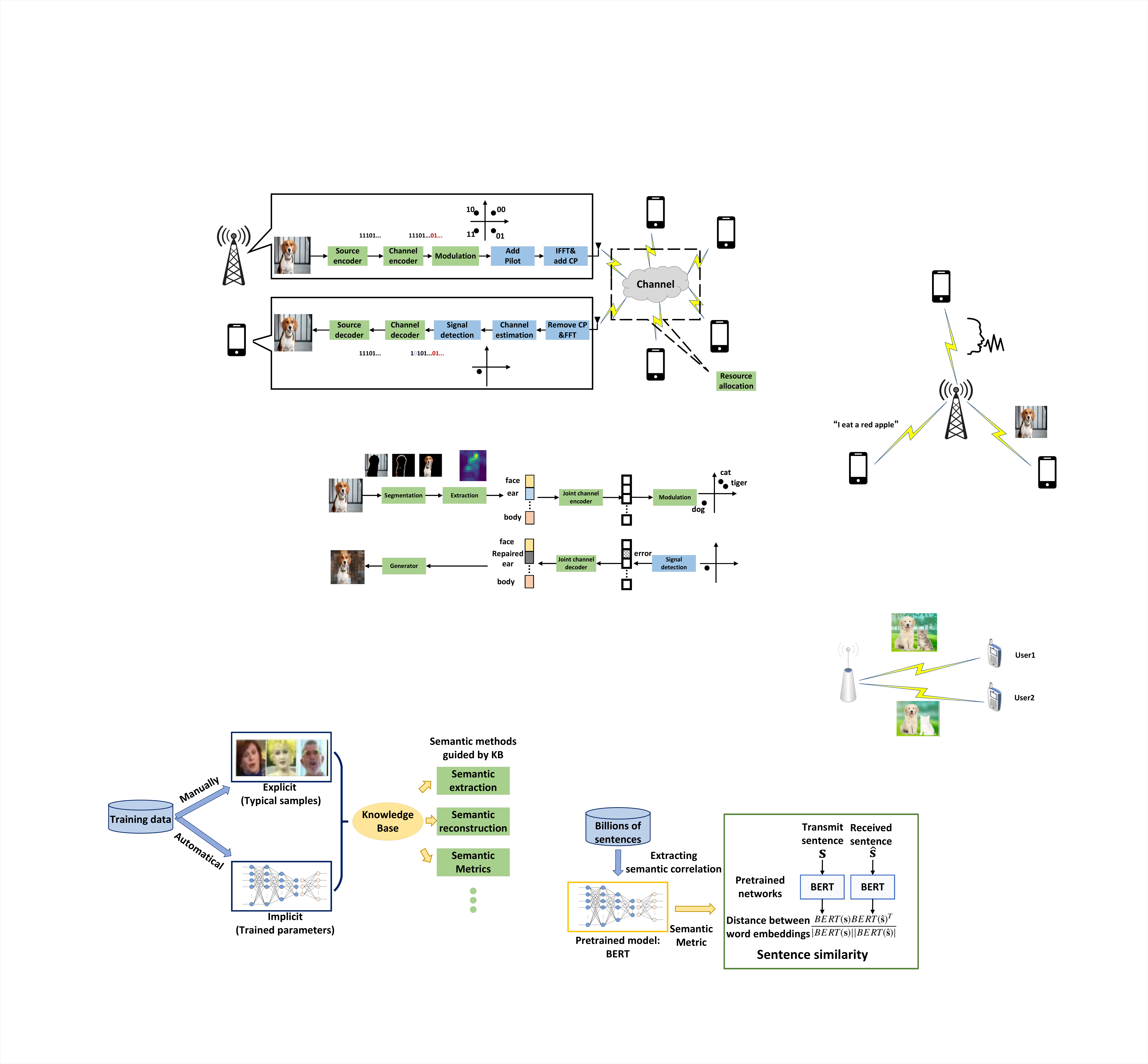}%}\\
	%	\subfloat[ ]{
	%		\includegraphics[width=4in]{figure//basic_nets}}
	
	% where an .eps filename suffix will be assumed under latex,
	% and a .pdf suffix will be assumed for pdflatex; or what has been declared
	% via \DeclareGraphicsExtensions.
	\caption{A conventional OFDM wireless communication systems. The green modules can be potentially redesigned for semantic communications while the blue modules are still conventional.  }
	\label{Conventional}
\end{figure*}

 	 Different from the existing survey or tutorial literature, such as \cite{gunduz2022beyond,lu2022rethinking,shi2021semantic}, this article focuses on wireless semantic transmission based on revising or redesigning the modules in the conventional communications. 
We first look at the  conventional modules as in a wireless communication system in Fig. \ref{Conventional} and then discuss its limitation. Then, the novel  changes to facilitate  semantic transmission are described. In general, semantic communications thoroughly reform the transmission paradigm as demonstrated by two examples in this article. Since the development of semantic communications is still in its infant phase, we highlight some challenges on  practical wireless semantic communications.

	The rest of this article is organized as follows. Section \uppercase\expandafter{\romannumeral2} introduces a conventional wireless communication system and indicates the growth trend of wireless terminals and multimodal requirements.   Section \uppercase\expandafter{\romannumeral3}  describes how semantic transmission affects the design of different modules in communication systems.   Section \uppercase\expandafter{\romannumeral4} presents two examples on semantic channel coding and modulation.   Section \uppercase\expandafter{\romannumeral5} provides some open issues on  wireless semantic communications.   Section \uppercase\expandafter{\romannumeral6} concludes this paper.

\section{Why wireless semantic communication?}
In this section, a conventional communication system is introduced first. Then, the new communication scenarios and requirements are discussed. Finally, the limitation of conventional modules are pointed out.
\subsection{A classic wireless communication system}

We use  orthogonal frequency division multiplexing (OFDM) as an example to discuss the difference between the conventional and the semantic communication systems since OFDM is widely used.
As shown in Fig. \ref{Conventional}, the source content, such as  images or texts, is compressed and converted into a bit sequence by a source encoder and then redundancy is added to the bit sequence by a channel encoder to cope with the channel distortion. Next, a proper modulation converts the bit sequence into a complex symbol sequence.  The pilot is inserted  for channel estimation before  inverse fast Fourier transform (IFFT). To facilitate OFDM demodulation, cyclic prefix (CP) is added before sending to wireless channels. For multi-user networks, limited wireless resources, such as bandwidth and transmission power, should be properly allocated to optimizing network performance.

Many modules at the receiver, such as demodulation, channel decoding, and source decoding, perform inverse operations of the corresponding modules at the transmitter. As shown in Fig. \ref{Conventional}, the FFT operation, channel estimation, and signal detection deal with the impact of channels. The errors in the detected bits are corrected by channel decoder and the source decoder restores the transmit content, such as image.

	\subsection{Growing demand of wireless services}
 Mobile work and online conferencing become  essential parts of our life, especially during the pandemic of COVID-19. For example, the transmission traffic has been increased over 60\% compared to that before the outbreak of COVID-19. In order to deal with the unbearable demand, some service providers, such as YouTube, can only reduce video qualities at peak times.   On the other hand, the users expect to enjoy a high-quality service, such as high-resolution videos, without  restrictions on time and place. As a result, semantic communication, which significantly improves transmission efficiency and  enhance user's experience, is desired.

Apart from improving the user's experience,  wireless networks  also  need to serve a huge number of terminals. For example,  autonomous cars rely on thousands of  sensors for data collection and cooperate  with other vehicles. The data transmission in vehicular networks usually serves specific tasks, where semantic communications are expected to play an important role.

 \subsection{Limitation of separate module design}
In the classic Shannon's paradigm, the channel coding has no need to consider the semantic meaning of the transmit content. Thus, the conventional modules follow the divide-and-conquer designs. However,  the code length is limited in low-delay scenarios, such as conferencing and autonomous drive.  On the other hand, the transmission features under a specific task has  strong correlation. Thus, focusing on bit-level transmission is not  efficient any more  in these situations. The content-related semantic methods are brought to the forefront. 

\section{Deep semantic system designs }
Most state-of-art works on semantic communication focus on the joint source-channel coding (JSCC) design.  Some methods redesign the modules in the conventional communication systems, such as modulation, signal detection, PAPR reduction, and resource allocation, for semantic communications. We should emphasize that all semantic systems are enabled by DL since it is still the only way to extract semantic meaning from the transmit content of the source.

\subsection{Knowledge base}
 \begin{figure}[!h]
	\centering

	\subfloat[ ]{
		\includegraphics[width=3.5in]{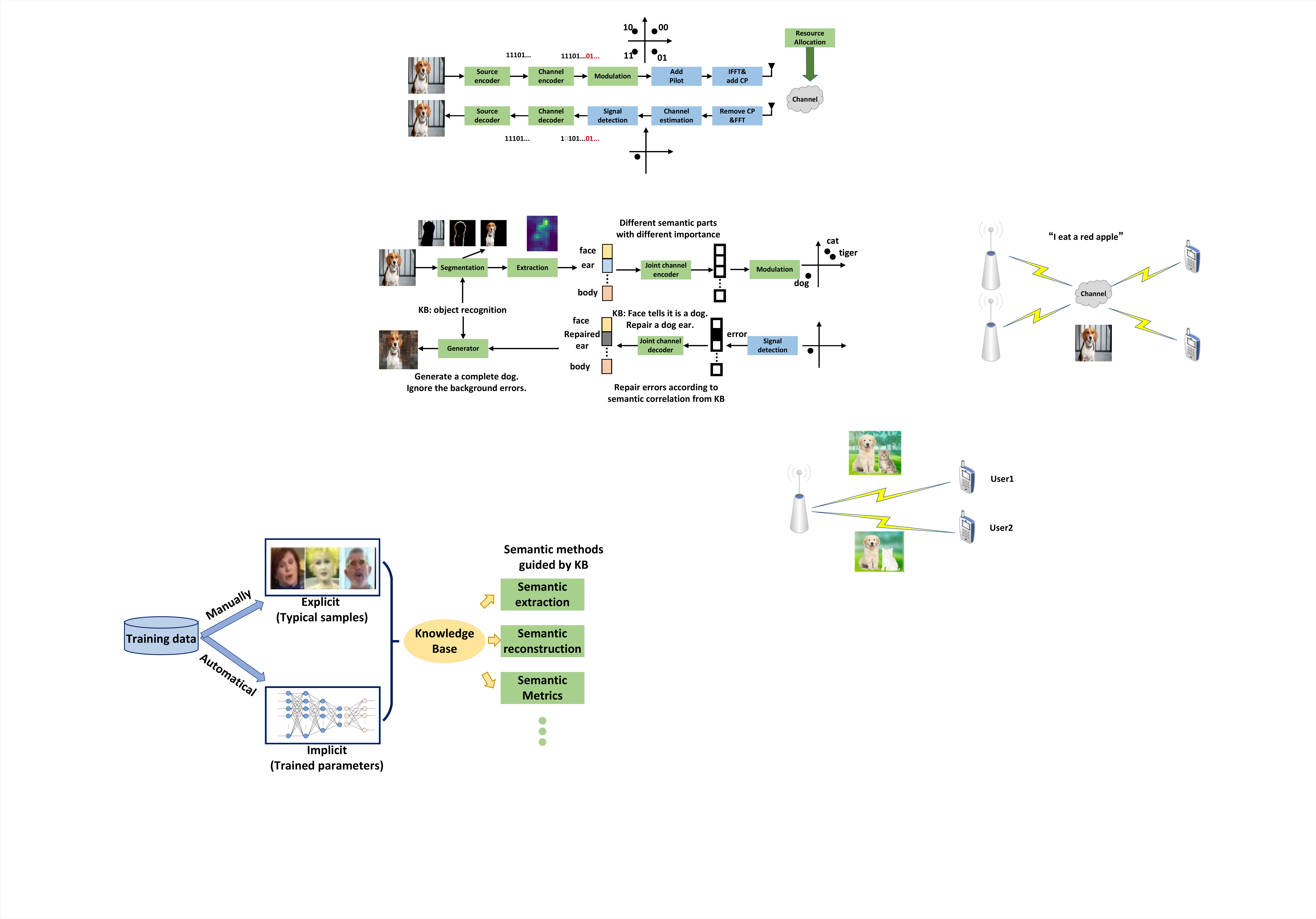}}\\
		\subfloat[ ]{
			\includegraphics[width=3.5in]{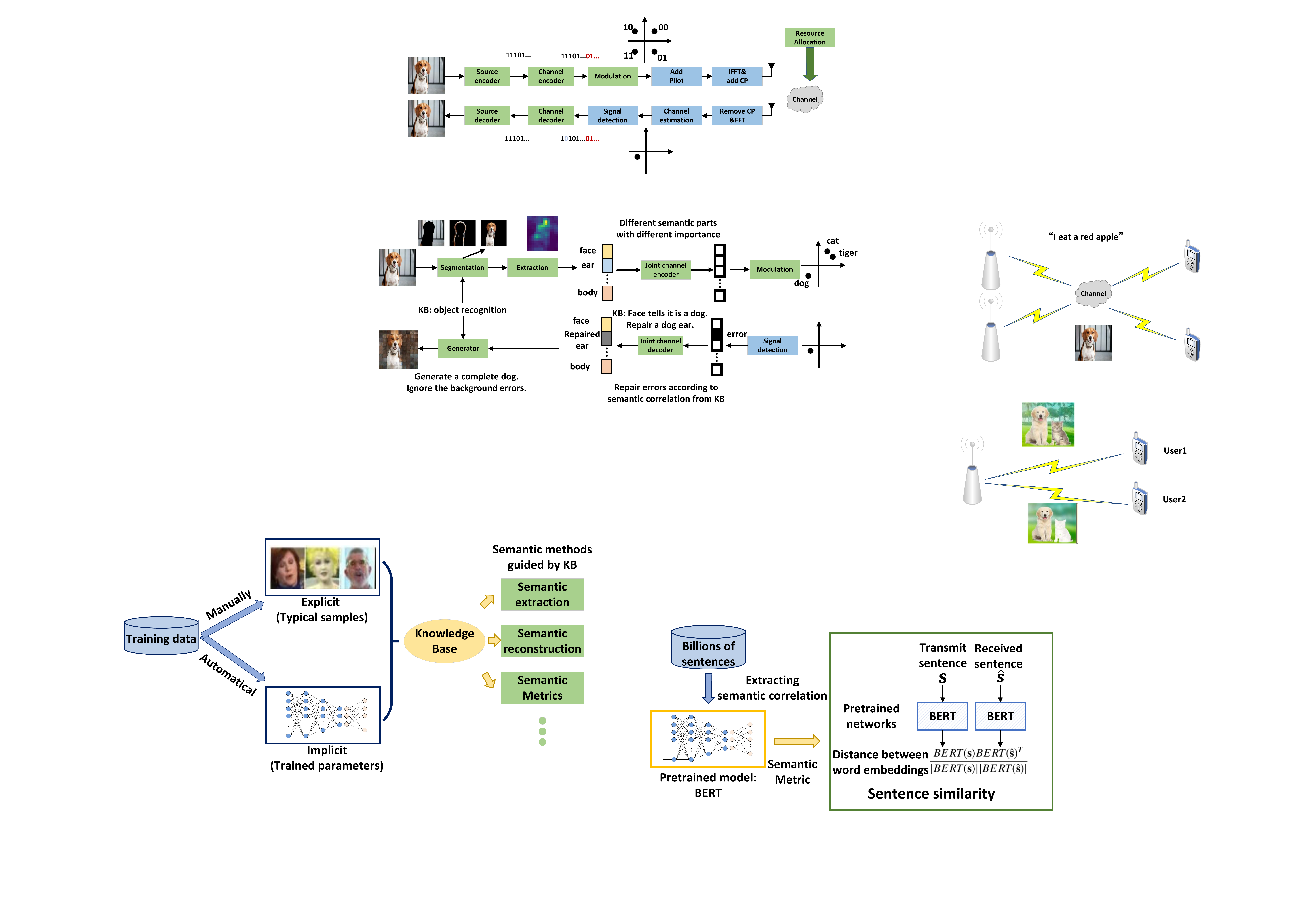}}
	
	% where an .eps filename suffix will be assumed under latex,
	% and a .pdf suffix will be assumed for pdflatex; or what has been declared
	% via \DeclareGraphicsExtensions.
	\caption{(a) Two representations of KB: explicit and implicit KBs. (b) An example of KB applications: a semantic metric, called sentence similarity, based on a pretrained BERT, which is trained under billions of sentences to establish an implicit KB. }
	\label{KB}
\end{figure}
To establish a semantic communication system, the KB plays a core role to guide the semantic extractor to only transmit the compressed unknown information and the semantic decoder to correct the  errors. In \cite{bao2011towards}, each transmitter and receiver pair have their local KB in addition to a KB shared by all transceivers when considering the consensus and disagreement of different equipment sets or users. In Fig. \ref{KB}(a), the local and shared KBs are categorized into implicit and explicit KBs. 

\begin{itemize}
		\item
	Explicit KB can be shared in specific tasks.  For example,  a talking-head video usually has a static background and a specific speaker. The photo of the speaker can be shared to the receiver as an explicit KB \cite{jiang2022wireless}, which contains the unchanged semantic information, such as the appearance of the speaker. This explicit KB can be replaced with a new photo easily once the speaker is changed. 
	
	\item 

Implicit KB is extracted automatically by DL methods and is represented by the trained parameters. Thus, implicit KB is naturally established after E2E training of the semantic transmitter and receiver. However, implicit KB is inexplicable  and inflexible. Retraining or transfer training is necessary to form a new KB when facing new semantic scenario.

\end{itemize}

After establishing a KB, many semantic-based operations can be designed, such as semantic extraction, reconstruction, and metric. The metric in Fig. \ref{KB}(b), called sentence similarity, is used in many studies, such as  \cite{guo2022signal,shao2022semantic,jiang2022deep,mu2022heterogeneous}. Sentence similarity is based on a pretrained model called BERT to extract knowledge from billions of sentences. After the sentence correlation is well-learned by BERT, the distance between the embedded word vectors can be considered as a measurement of sentence similarity. The value of sentence similarity reaches its maximum, 1, if the two sentences are exactly same.

In fact, most of the semantic features are difficult to be represented explicitly and rely on the development of DL. The attention-based DL is brought to forefront because the corresponding networks can distinguish the importance of different source parts.

\subsection{Source coding based on semantic segmentation and extraction}
 \begin{figure*}[!h]
	\centering

%	\subfloat[ ]{
		\includegraphics[width=6.5in]{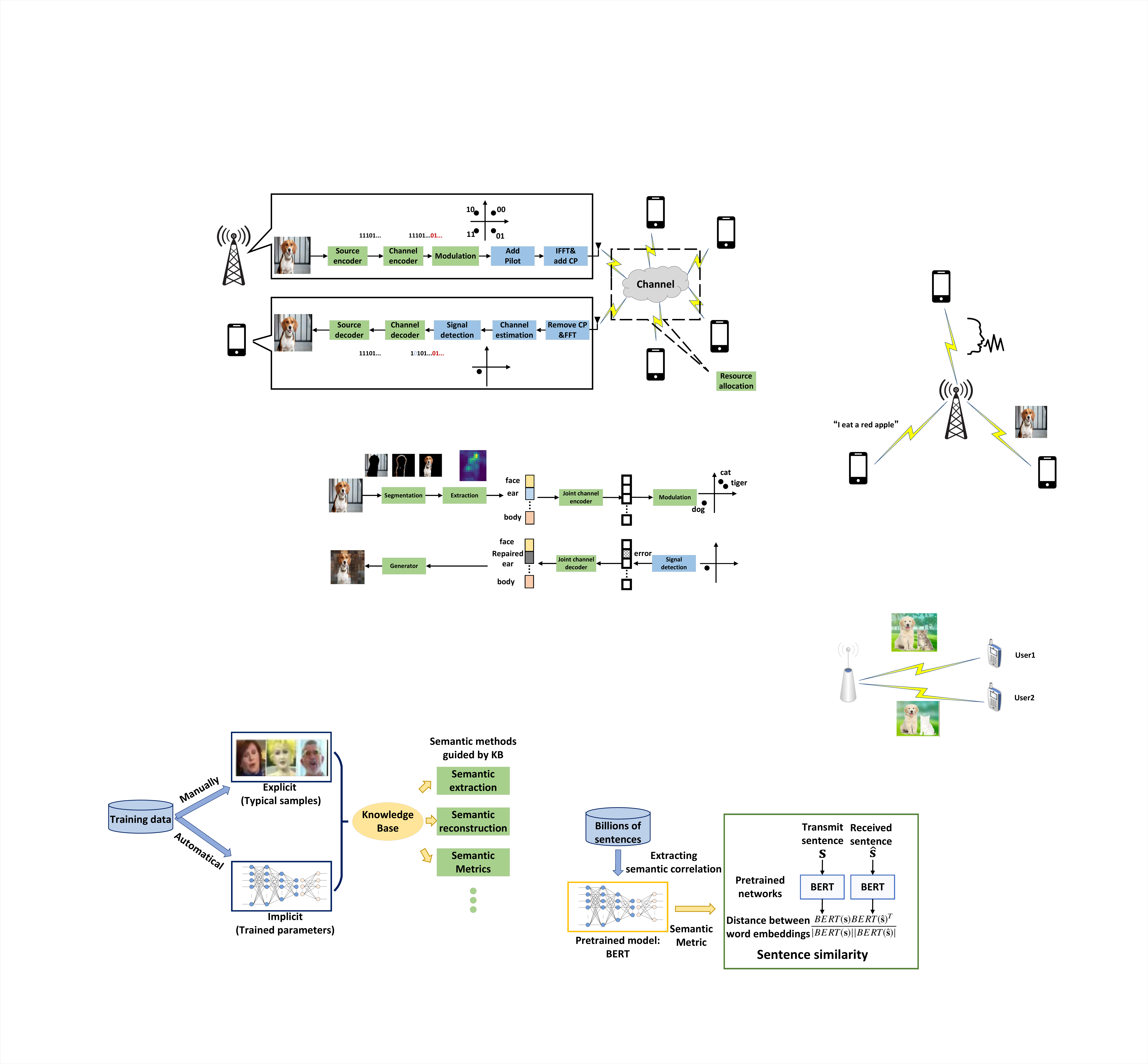}%}\\
	%	\subfloat[ ]{
	%		\includegraphics[width=4in]{figure//basic_nets}}
	
	% where an .eps filename suffix will be assumed under latex,
	% and a .pdf suffix will be assumed for pdflatex; or what has been declared
	% via \DeclareGraphicsExtensions.
	\caption{ Examples of the source-channel coding and the modulation based on the semantic extraction, segmentation, and metrics, where the green modules represent the semantic methods. }
	\label{Semantic_modules}
\end{figure*}
Source coding aims to reduce the transmission payload by compressing the original source content. Based on the existing semantic segmentation and extraction methods, the source coding can be applied efficiently. In Fig. \ref{Semantic_modules}, the capabilities of semantic segmentation and extraction are illustrated.

Semantic segmentation divides the source into different semantic parts and each is usually in different degrees of importance. The segmentation relies on the KB in the specific scenario. A sentence may be divided into nouns, verbs, adjectives, and adverbs, where nouns and verbs are usually more important than adjectives and adverbs. An image may be divided into the background, sketches, and objects, where the background usually attracts less than the objects. After semantic segmentation,  different parts of the source are protected with differently degrees and the most important semantic part is protected with the best channel condition.

Semantic extraction reduces the redundancy of the source based on the shared KB. Because some semantic information can be restored directly by the shared KB, only the key semantic information is required to transmit. With the knowledge of the appearance, only some points \cite{jiang2022wireless} representing the facial expression are transmitted to restore a talking-head.

At the receiver, the source decoder combines all semantic parts of the source and the semantic information from the shared KB and then generates the required content. Restoring the source completely is a conventional goal without any KB about the task at the receiver. If the requirement of the receiver is an object recognition task, the objects and sketches are prior are more important than the background when transmission resources are not sufficient. 

Since semantic segmentation and extraction are based on the KB that is shared by transmitter and receiver, the transmission payload can be significantly reduced compared to the conventional source coding methods, which only compress the source in bit-level. 

\subsection{Channel coding based on joint design and training}

The compressed source can be  protected by  conventional channel coding and decoding, such as Reed-Solomon (RS) code in \cite{jiang2022deep}. However, JSCC, with the help of KB, can significantly improve error correction capability. The joint design  can automatically protect  different parts of the source with different code rates \cite{zhou2022adaptive}. When the transmission bandwidth is limited, the channel coding pays attention to some important semantic features  and  reduces the performance loss in semantic similarity. 

 Fig. \ref{Semantic_modules} shows the mechanism of channel coding. The channel encoder has the importance knowledge of  different parts of a dog, such as the face, ear, and body and protects them in different degrees. If the received codeword corresponding the ear in the dog's image is with errors, the conventional channel decoder corrects the errors only exploiting the redundancy but has no knowledge of the content. The semantic channel decoder can further correct the errors with the KB and restore the dog's ear if the dog is recognized accurately with other parts. 

Overall, the semantic-based channel coding has more ways to protect the transmit content and correct the transmission errors with the equipment of the KB. Thus, the semantic coding can still protect some semantic features when the channel environment is nasty and the transmission errors surpass the correction capability of the conventional channel coding.

\subsection{Physical modules based on minimizing semantic errors}

The conventional physical modules  are usually optimized independently.  For example, the modulation and signal detection are designed to minimize the bit-error rate (BER).  The channel estimation is minimizing the mean-squared error (MSE) between the estimated and true channels.  For the semantic communication, the transmit features have different importance levels and different features are with semantic correlation, which can be exploited for semantic recovery.

\textbf{Modulation:} The modulation in wireless semantic communication needs to be reconsidered due to the new performance metrics. The conventional modulation methods, such as quadrature amplitude modulation (QAM), minimize BER but are content-unaware. In fact, the bits or transmit features are not equally important. For example, the symbols representing a tiger are close to the those representing a cat rather than a dog because some parts of a cat is similar to those of a tiger and has no much impact if they are replaced mistakenly. For semantic text transmission in \cite{guo2022signal}, the sentence similarity is exploited to form a new modulation method.

\textbf{PAPR:} PAPR is  also an issue for a practical system because  high PAPR challenges the hardware devices. In \cite{shao2022semantic}, the PAPR is considered as an extra loss function and the semantic network is trained to minimize the semantic and PAPR losses together, which  means PAPR reduction and  semantic performance metric are balanced.

\textbf{HARQ:} Under varying channel environments, the HARQ with acknowledgment (ACK) feedback is essential for a successful transmission. The conventional HARQ is based on a channel coding for forward error correction (FEC) and a cyclic redundancy check (CRC) code for error detection. All these methods are designed in bit-level. In \cite{jiang2022deep}, the semantic-based JSCC  is introduced to replace channel coding and the transmission errors can be corrected according semantic correlation. Besides, the sentence similarity is used to replace the CRC methods and the received sentences with similar meaning can be accepted with retransmission. In this way, the  performance and throughput are significantly improved.   

\textbf{Channel state information (CSI) feedback:} Data hiding is also a potential DL method for semantic transmission and is exploited to remove transmission payload in CSI feedback \cite{guo2022eliminating}. Besides, CSI feedback can help the semantic channel coding\cite{jiang2022wireless} to assign important information to the subchannels with high signal-to-noise ratios (SNRs).

\subsection{Resource allocation for semantic requirements of different users}

 \begin{figure}[!h]
	\centering

%	\subfloat[ ]{
		\includegraphics[width=3in]{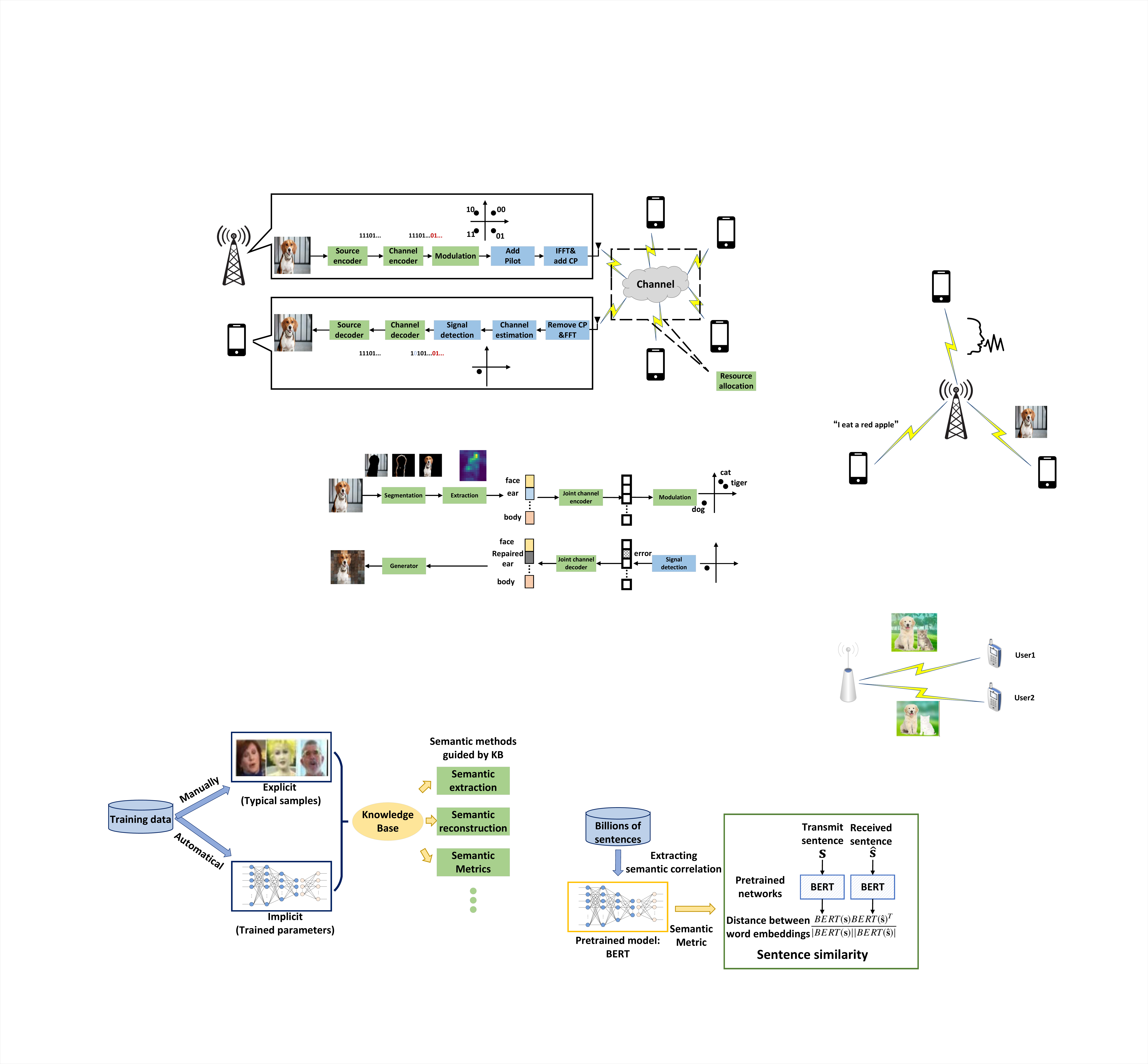}%}\\
	%	\subfloat[ ]{
	%		\includegraphics[width=4in]{figure//basic_nets}}
	
	% where an .eps filename suffix will be assumed under latex,
	% and a .pdf suffix will be assumed for pdflatex; or what has been declared
	% via \DeclareGraphicsExtensions.
	\caption{ Resource allocation for different users transmitting multimodal information, such as text, image, and speech.}
	\label{RA}
\end{figure}

	Resource allocation is increasingly crucial due to the growth of users and terminals. With a proper allocation method, the quality of service for all users is guaranteed and the transmission resources can be saved. The conventional goal of the resource allocation is transmitting the required bits within the required time for a user. Now that the number of bit errors is replaced by the semantic performance metrics, such as sentence similarity, in semantic communications, the resource allocation method should changed accordingly.
	
	Resource allocation in semantic communication has been initially studied  in \cite{yan2022resource}. The performance metric is changed to sentence similarity and the success probability of tasks is maximized. The successful transmission requires a proper sentence similarity; otherwise, the meaning may be misunderstood.   
	
	However, the semantic performance metrics vary with the contents and tasks. Thus, the resource allocation in semantic systems is heterogeneous and the complexity is sharply increased.  As shown in Fig. \ref{RA}, text, image, and speech are transmitted to three users at the same time. The semantic parts in these contents are with different importance levels. When transmission resources are limited, the most important semantic requirements should be satisfied, i. e., the meaning of text and speech should be unchanged and  the main object in the image should be distinct. When sufficient transmission resources are available, these resources are allocated for a high-quality service, i. e.,  the sentence and the speech are accurate and all pixels of the image are restored. Thus, semantic resource allocation should distinguish the importance levels of different transmitted  parts and adaptively allocate the resource according to channel conditions and user's requirements.

\section{Two examples on wireless semantic communications}

We demonstrate the superiority and the changes brought by wireless semantic communications. Since the error correction capability plays a key role, especially when the channel environment is poor, two examples about channel coding and modulation will be presented in this section.

\subsection{Rate of sentence success transmission under varying channels}

 \begin{figure}[!h]
	\centering

	%	\subfloat[ ]{
	\includegraphics[width=3in]{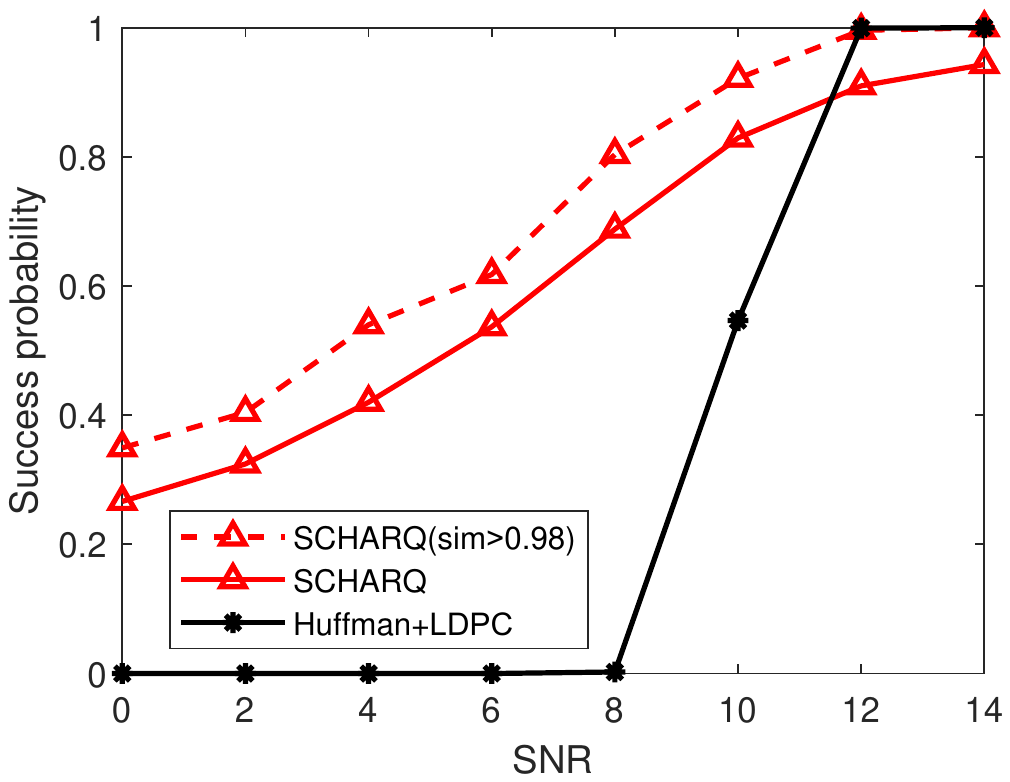}%}\\
	%	\subfloat[ ]{
	%		\includegraphics[width=4in]{figure//basic_nets}}
	
	% where an .eps filename suffix will be assumed under latex,
	% and a .pdf suffix will be assumed for pdflatex; or what has been declared
	% via \DeclareGraphicsExtensions.
	\caption{ The rate of success transmission for semantic-based and conventional Huffman+LDPC-based HARQ methods. SCHARQ (sim$>$0.98) means the successful transmission only requires sentence similarity to be larger than 0.98.}
	\label{su}
\end{figure}

The conventional HARQ uses a forward error code to correct the transmission errors and a CRC code detects the errors. However, this method only concentrates on the bit-level. The semantic-based coding method learns to correct the sentence with the semantic correlation. Semantic-based error detection finds the sentences with the changed meaning and the sentences with unchanged meaning has no need to retransmit. Thus, the success transmission rate can be improved with the novel FEC and error detection methods. The detailed implementation is shown below.

\textbf{Input sentence:} The sentences in the European Parliament are chosen with lengths between 4 and 30 words. There are 100,000 sentences for training and 10,000 sentences for testing.

\textbf{Huffman+LDPC-based HARQ:} The input sentences are compressed by Huffman coding and then protected by low-density parity-check (LDPC) coding. The  LDPC-based HARQ can be easily realized by puncturing LPDC codeword into different code rates. After Huffman coding, the average bits of a sentence is about 460 bits. The max code rate of LDPC codeword is 5/16 and the max code length is about 1,470 bits. 

\textbf{SCHARQ:} The semantic coding (SC)-based incremental redundancy HARQ method has been proposed in \cite{jiang2022deep}. The input sentence is encoded into different codewords by a Transformer-based JSCC to extract the semantic correlation. The error detection method can also be replaced by a sentence similarity.  According to the ACK feedback, different numbers of codewords are transmitted, which means the code rate is adaptive. The maximum code length is 1,000 bits for a single sentence.

\textbf{Experimental setup:} These two methods are implemented into an  OFDM system with 16-QAM modulation. Each block has eight OFDM symbols and each OFDM symbol has 64 subcarriers. The first OFDM is used for pilot and others for  data.   For  time-varying channels, the rate of success transmission is calculated by transmitting 10,000 sentences.

From Fig. \ref{su}, SCHARQ significantly improves the number of correct received sentences under time-varying channels when SNR is low, where the conventional Huffman+LDPC method cannot correct the received sentences because the number of errors is over the correction capability. However, SCHARQ has worse successful transmission probability when SNR is larger than 12 dB. That means the network always has a little performance degradation due to the difference between training and testing data. SCHARQ(sim$>$0.98) means the transmission can be regarded as a success when the sentence similarity is larger than 0.98. The value of 0.98 is so close to 1, and therefore, the sentence meaning has no change. In this way, SCHARQ(sim$>$0.98) always transmits more sentences than those of SCHARQ and Huffman+LDPC.

 \begin{figure}[!h]
	\centering

	\subfloat[ ]{
		\includegraphics[width=3in]{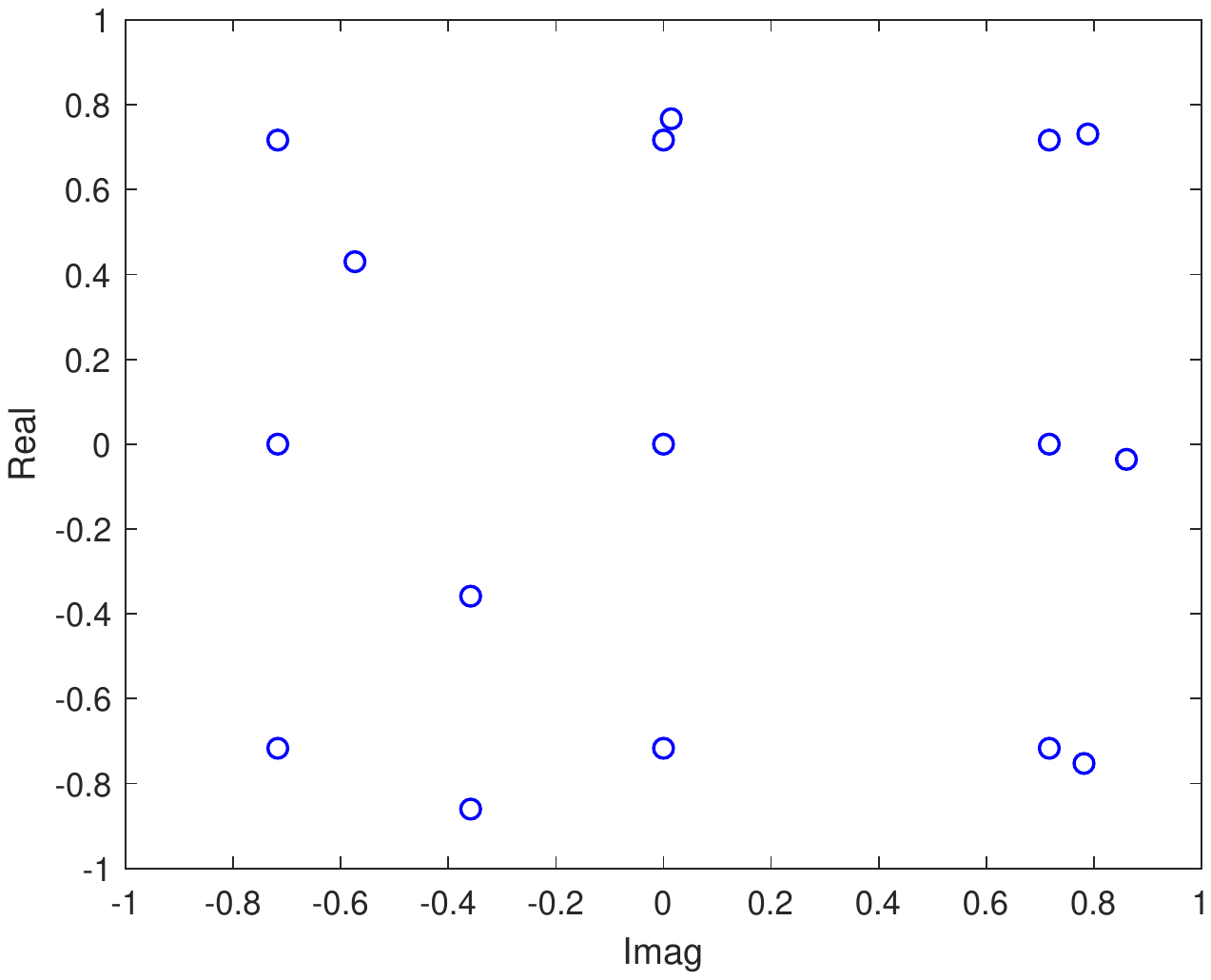}}\\
	\subfloat[ ]{
		\includegraphics[width=3.1in]{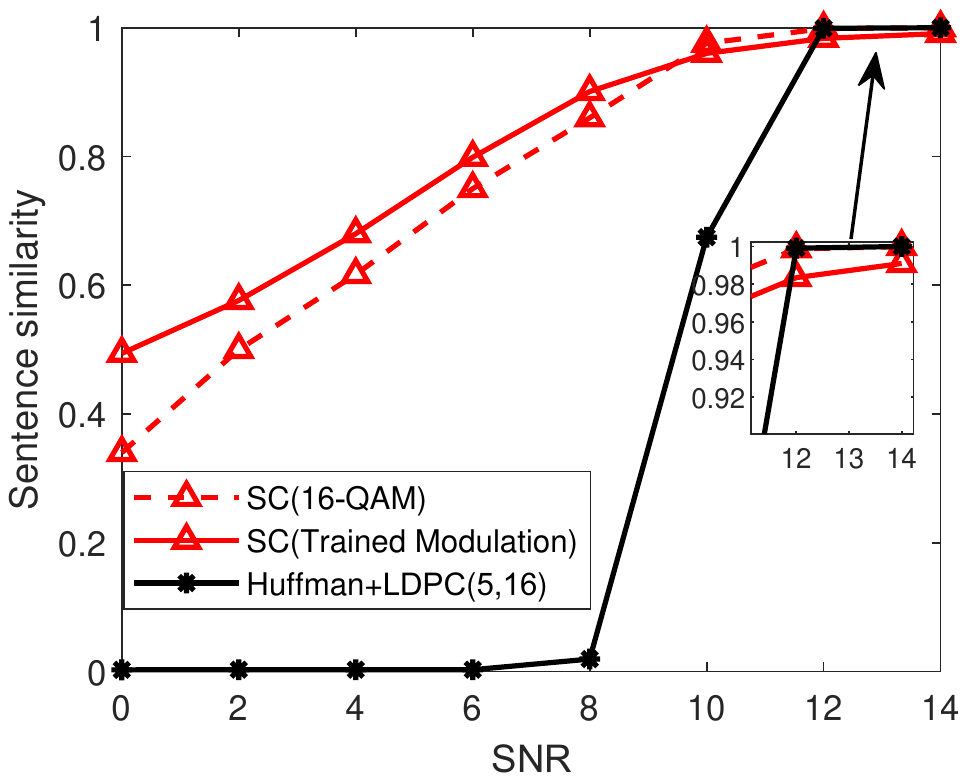}}
	
	% where an .eps filename suffix will be assumed under latex,
	% and a .pdf suffix will be assumed for pdflatex; or what has been declared
	% via \DeclareGraphicsExtensions.
	\caption{ (a) The learned 16 possible locations of the constellation points trained to maximize sentence similarity and these locations are not distributed uniformly.  (b) Sentence similarity performance of semantic methods with 16-QAM and trained modulation.}
	\label{Modu}
\end{figure}
\subsection{Modulation for semantic transmission}
 The conventional designed constellation points are usually distributed uniformly like QAM and phase shift keying (PSK) because all the bits are equivalent. In this paradigm, an E2E semantic JSCC is established based on \cite{xie2020deep} and  each encoded symbol is  quantified into 4 bits, which are converted into 2 real numbers by a dense layer with tanh activation function. The 2 real numbers denote a two-dimensional coordinate, which means 4 bits are mapped to a constellation point with 16 locations. Each sentence is modulated into 80 constellation points by E2E training and  the loss function is replaced by maximizing  sentence similarity. Therefore, two similar sentences may be mapped into constellation points with a close distance.

As shown in Fig. \ref{Modu}(a), the trained constellation points  are different from those of the conventional QAM method. The distance between two adjacent points varies because these constellation points are unequal in sentence similarity. If a point  mistakenly detected as another has no effect on the sentence meaning, their distance is close. 

In Fig. \ref{Modu}(b), the trained modulation method has better performance in sentence similarity than 16-QAM, especially when SNR is low. However, the trained modulation is little worse than 16-QAM when SNR is higher than 15 dB  because 16-QAM has nearly no detection error when SNR is high. In contrast, some trained modulation points are too close to others and still cannot be correctly detected when SNR is between 15 and 20 dB. 

From above examples, the semantic methods bring a novel way to transmit the meaning of content. The changed performance metric focuses on bringing an unchanged semantic information rather than reducing bit errors. Thus, the transmission modules are also redesigned to protect the semantic features. The semantic methods can  cope with terrible channels and limited bandwidth.

\section{Open issues}
Semantic communication achieves great success recently but brings new issues. Some novel transmission methods rely on the KB and E2E training, which is inexplicable  and inflexible. Moreover, the  new metrics in semantic communications also make the design complex. Thus, further work is still needed to establish a practical wireless semantic communication system. 

\subsection{ KB establishment and update}
As an essential part of  semantic communication, a proper KB directly affects the transmission accuracy. However, a major part of a KB is implicitly extracted in the trained parameters in neural networks. These parameters can enhance the semantic modules under the content with the same domain knowledge as the training data but mislead the system facing a new content. This challenge is inevitable when the bit transmission is replaced with the content-related transmission. 

Establishing a KB is the initial step and the requirements of the application scenario should be considered.  In a common video transmission, the semantic segmentation divides the current video frame into different semantic bodies but all these bodies are needed to transmit. In a specific video conferencing (a talking-head video), the background is static and the speaker is known. Thus, only some points representing the changing expression are transmitted and other parts are generated from the shared KB. The semantic communication system for video conferencing  reduces resources significantly but cannot be applied in common video transmission. 

Updating a KB is also important because offline training cannot cover all semantic scenarios while online data is hard to collect. Thus, the few-shot training, such as domain adaption, is helpful. Federated learning is also a potential direction to share KB for different terminals. 

\subsection{Varying wireless channels}
Varying wireless channels are rarely studied in E2E training manner because the multiplicative wireless channels prevent from the gradient passing. Thus, the additive white Gaussian noise (AWGN)  or binary symmetric channels (BSC) are widely used when testing the designed systems. However, wireless channels are time-varying and frequency-selective. Therefore,  they have different gains at different subchannels or times.   Furthermore, the features of wireless channels can be exploited by semantic communication methods, such as transmitting the important semantic features in the subchannels with high SNRs. 

Meanwhile, the fixed  architecture is  unsuitable for time-varying wireless channels. After training, each transmit symbols carry a constant amount of semantic information and cannot adapt to the changing channel conditions.

\subsection{Multimodal sources and requirements}
In conventional communication systems, all kinds of sources are converted into bits and their correlation is ignored. The requirement of the transceiver is also simplified into reducing bit errors. In contrast, the multimodal sources and requirements are sensed by semantic communication systems and  bring challenges. 

Because there is no unified metric for multiple sources and requirements, different contents and tasks are designed and optimized independently but transmitted in the same system. In \cite{zhang2022unified}, the codewords are divided into shared  and private features. Specifically, each task has its private features and reaches a good performance under a unified transmission model. Apart from a unified transmitter for different requirements, the source allocation for multiple sources and requirements is also an issue in multi-user systems. The semantic features from different users are  in different modes and the varying tasks require the resource allocation methods being more adaptive with changing KBs.

\section{Conclusions}
In this article, we have first discussed the limitation of the modules in the conventional communications and the new demands of semantic communications. With the increasing of wireless terminals and multimodal requirements, semantic communication is more and more desired to reduce the transmission payload through focusing on the important semantic information by sharing the KB in advance. After introducing the concept of semantic communications, we have investigated how to revise or redesign some modules in the conventional communication. Using two examples, we have shown that  the semantic-based JSCC and modulation are much better than the conventional methods. Finally, the open issues have been provided.

	\bibliographystyle{IEEEtran}
	\bibliography{bibtex0320}
	
	
	%%%%\begin{IEEEbiography}{Michael Shell}
	%%%%Biography text here.
	%%%%\end{IEEEbiography}
	%%%%
	%%%%% if you will not have a photo at all:
	%%%%\begin{IEEEbiographynophoto}{John Doe}
	%%%%Biography text here.
	%%%%\end{IEEEbiographynophoto}
	%%%%
	%%%%% insert where needed to balance the two columns on the last page with
	%%%%% biographies
	%%%%%\newpage
	%%%%
	%%%%\begin{IEEEbiographynophoto}{Jane Doe}
	%%%%Biography text here.
	%%%%\end{IEEEbiographynophoto}
	
	
	%\vfill
	

\end{document}